\documentclass[a4paper,10pt]{article}

\usepackage{latexsym}   \usepackage{amssymb}
\usepackage{amsmath}    \usepackage[mathscr]{eucal}
\usepackage{cite}

\newcommand{\ncm}[2]{\newcommand{#1}{#2}}
\ncm{\dss}{\displaystyle}
\ncm{\tss}{\textstyle}

\ncm{\be} {\begin{equation}}
\ncm{\ee} {\end{equation}}
\ncm{\bl} {\begin{eqnarray}\begin{array}}
\ncm{\el} {\end{array}\end{eqnarray}}
\ncm{\ba} {$$\begin{array}}
\ncm{\ea}{\end{array}$$}
\ncm{\bea}{\begin{eqnarray}}
\ncm{\eea}{\end{eqnarray}}

\ncm{\He} {\hbox{}{$^3$He}\hbox{ }}
\ncm{\Hep}{\hbox{}{$^3$He}}
\ncm{\Ht} {\hbox{}{$^3$H}\hbox{ }}
\ncm{\Htp}{\hbox{}{$^3$H}}

\ncm{\OH} {{\displaystyle {\frac{1}{2}}}}
\ncm{\Oh} {{\textstyle    {\frac{1}{2}}}}
\ncm{\oh} {{\scriptstyle  {\frac{1}{2}}}}
\ncm{\nh} {{              {\frac{1}{2}}}}
\ncm{\tH} {{\textstyle    {\frac{3}{2}}}}
\ncm{\ths}{{\scriptstyle  {\frac{3}{2}}}}
\ncm{\ott}{{\textstyle    {\frac{1}{3}}}}
\ncm{\oTt}{{\textstyle    {\frac{2}{3}}}}

\ncm{\vp}    {\vec p}
\ncm{\vq}    {\vec q}
\ncm{\vpqT}    {( \vec p, \vec q;  \, T  ) }

\ncm{\vph}   { \overset{\:{\scriptscriptstyle\land}}{\vec p} }
\ncm{\vqh}   { \overset{\:{\scriptscriptstyle\land}}{\vec q} }
\ncm{\vpsvq} { \vph\:\cdotp\vqh }
\ncm{\vnh}   { \overset{\:{\scriptscriptstyle\land}}{\vec n} }

\ncm{\Ts} {\mathscr{T}}
\ncm{\Ls} {\mathscr{L}}
\ncm{\Ss} {\mathscr{S}}

\ncm{\Pinv} {\mathcal{P}}  
\ncm{\Trev} {\mathcal{T}}  

\ncm{\Spr} {\mbox{Sp}\;}

\title{            Tensor and Operator Forms \\
                   of {${^3}$He} and {${^3}$H} Wave Functions  \\
                   for Parity--Violating Nuclear Forces}
\author{
V.~Kotlyar\thanks{\textit{E-mail address:} kotlyarv@kipt.kharkov.ua} \\
{\normalsize National Science Center "Kharkov Institute of Physics and Technology",} \\
{\normalsize 61108 Kharkov, Ukraine }}

\begin{document}

\maketitle
\begin{abstract}
Tensor representation (TR) for wave function (WF) of three--nucleon bound state
with the total angular momentum $I=1/2$ is discussed.
The WF in TR has 16 complex components depending on vectors of relative momenta.
Constraints on the WF imposed by requirements of invariance with respect to
space inversion and time reversal are studied.
Both parity--even and parity--odd components of the 3N bound state
are constructed using 16 scalar functions.
The arguments of the functions are
magnitudes of relative momenta and scalar product of the momenta.
With nuclear forces being time--reversal invariant these functions are real.
The WF in TR is converted into an operator form (OF),
accounting for parity violating contributions.
Properties of operator representations for WFs of 2N and 3N nuclei are compared.
\\[0pt]
\\[0pt]
\end{abstract}

\section{Introduction}

Structure and general properties of the three-- and four--nucleon bound state WFs
were investigated by E.Gerjuoy and J.Schwinger~\cite{GerjuoySchwinger} under the assumption
that the nuclear forces are invariant with respect to the space inversion.
Three--nucleon system of even parity that has the total angular momentum $I=\Oh$
was considered in the paper.
It was shown that the tritium WF can be obtained
from the 3N spin state with zero spin in two--nucleon subsystem
making use of a set of spin--angular operators multiplied by scalar functions.
The operators are constructed from vectors of the Jacobi coordinates and
Pauli matrices of nucleons.
The scalar functions depend on the magnitudes of the Jacobi coordinates and
cosine of the angle between them.
All three-nucleon states
$^{2\Ss+1}\Ls_I={^2S}_\oh,$ $^2P_\oh,$ $^4P_\oh,$ and $^4D_\oh$
with total orbital momentum $\Ls=0,1,2$ and total spin $\Ss=\Oh,\tH$
that may appear in 3N bound state with $I=\Oh$
are generated by the operator set proposed in~\cite{GerjuoySchwinger}.

Relationship between the OF and
the partial--wave decompositions~\cite{Glockle83} of the WF
was investigated in~\cite{Fachruddin04}. Eight scalar functions,
needed to build up the WF, were calculated for modern realistic models of nuclear forces.
Results of Ref.~\cite{Fachruddin04} allow one to compute WFs of \He and \Ht nuclei
in momentum space\footnote{The functions can be downloaded from \\
http://www.phy.ohiou.edu/$\sim$elster/h3wave/index.html}.

An OF for the WF was derived in~\cite{Kotlyar05} transforming WF~\cite{Kotlyar87} in TR,
in which the WF depends on vectors of relative momenta $\vp$ and $\vq.$
Operators~\cite{Kotlyar05} and corresponding scalar functions
$\psi_\lambda(p,q,{\vp\cdot\vq}/(pq)),$ where $\lambda=1,\dots,8,$
turned out to be superpositions of ones employed in~\cite{Fachruddin04}.
WF in the TR~\cite{Kotlyar87} was used in~\cite{Kotlyar05}
to calculate and analyze structure of spin--dependent momentum distributions
of nucleons and proton-deuteron clusters in 3N nuclei.
Earlier~(see, e.g.~\cite{Kotlyar87,KotlyarMelnikShebeko95}) one-- and two--nucleon mechanisms
of $\gamma{^3\text{He}}\to\text{pd}$ were studied with WF of \He in~TR.

Parity--conserving nuclear forces and components of WFs with positive parity
were discussed in Refs.~\cite{GerjuoySchwinger,Fachruddin04,Kotlyar05}.
In this paper we study structure of \He and \Ht WFs in tensor and operator representations
for parity violating interaction between nucleons.

In Sect.~2 properties of the WF in TR
that can be inferred using space inversion and time reversal are discussed.
We derive constraints imposed on components of the tensor WF
by requirement of time--reversal invariance in the case parity conservation does not hold.
The corresponding linear relations restrict the number of independent scalar functions
$\psi_\lambda$
needed to construct the WF in TR
and limit the amount of the terms in operator representation for the WF.
In Sect.~3 we demonstrate that both parity-even and parity--odd components of the WF in TR
can be built up using real scalar functions $\psi_\lambda$
with $\lambda=1,\dots,16.$
OF for the WF is derived in Sect.~4.
Common features in structure of the tensor and operator representations for
the 2N and 3N bound states are considered in Sect.~5.
In Appendix~A we treat the constraints placed on components of the tensor WF
by requirement of time--reversal invariance
in the degenerate case of collinear relative momenta.
Scalar functions $\psi_\lambda$
are expressed through components of the tensor WF in Appendix~B.


\section{Tensor Form of the 3N Bound--State Wave Function}\label{WFTensorForm}

In TR~\cite{Kotlyar87} WF of the three--nucleon bound states
with the total angular momentum $I=\Oh$ and its projection $m'$ reads
\be\label{PsiVec}
\Psi^{SMm}_{m'}(\vp,\vq\, )={_{23,1}\!\left<\vphantom{\Psi;\Oh m'}
\vp\:\vq; SM\Oh m\right|}\left.\Psi;\Oh m'\right>. \ee
Indices $23,1$ indicate the choice of the Jacobi momenta and
the order nucleon spins are coupled.

Eq.~\eqref{PsiVec} is written in the center of mass system of the nucleus.
The Jacobi momenta are
\be\label{JacobiMom}
\vp=\Oh(\vec k_2-\vec k_3), \quad \vq=
{\textstyle {\frac{1}{3}}}(2\vec k_1-\vec k_2-\vec k_3), \ee
with $\vec k_n$ being the momentum of the $n${\it th} nucleon $(n=1,2,3).$
The total spin of nucleons 2 and 3 is $S.$
The two--nucleon spin states are defined as
\be\label{SMSstate23}
\left|SM_S\right>\equiv\left|SM_S\right>_{23}=
\sum_{m_2,m_3} C^{SM_S}_{\oh m_2 \oh m_3} \left|\Oh m_2\right>_2 \left|\Oh m_3\right>_3, \ee
where $C^{a\alpha}_{b\beta c\gamma}$
is the Clebsch--Gordan coefficient~\cite{Varshalovich}.
Vector $\left|\Oh m_n\right>_{\!n}$ is an eigenstate of
the nucleon spin operator squared $\vec{s}^{\:2}(n),$
which belongs to the eigenvalue $m_n$ of $s_z(n).$

The isospin formalism is not employed in the present paper.
The nucleon labeled by 1 is chosen to be neutron (proton) for \He (\Htp) nucleus.
In turn, nucleons 2 and 3 are protons (neutrons).
WF~\eqref{PsiVec} transforms under the permutation (2,3) of identical nucleons 2 and 3
according to
\be\label{PsiVec23transp} \Psi^{SMm}_{m'}(\vp,\vq\,)=(-1)^S\Psi^{SMm}_{m'}(-\vp,\vq\,). \ee

For a parity violating interaction between nucleons
nuclear states are superpositions of opposite parity terms
\be\label{PsiDefParityDecomp}
\left| \Psi;\Oh m'\right>=\sum_{N=0,1} \left|\Psi;\Oh m'; N\right>, \ee
where
$\Pinv\left|\Psi;\Oh m'; N\right>=(-1)^N \left|\Psi;\Oh m'; N\right>\!,$
and $\Pinv$ is space inversion. Components
\be\label{PsiVecDefParity}
\Psi^{SMm}_{m'}(\vp,\vq; N)={_{23,1}\!\left<\vphantom{\Psi;\Oh m'}
\vp\:\vq; SM\Oh m\right|}\left.\Psi;\Oh m'; N\right> \ee
of WF~\eqref{PsiVec}, which have definite parity $(-1)^N,$ obey
\be\label{PsiVecDefParityTransf}
\Psi^{SMm}_{m'}(\vp,\vq; N)=(-1)^N \Psi^{SMm}_{m'}(-\vp,-\vq; N). \ee

A consequence of~\eqref{PsiVec23transp} and~\eqref{PsiVecDefParityTransf} is
\be\label{PsiVec23ParityTransf}
\Psi^{SMm}_{m'}(\vp,\vq; N)=(-1)^{S+N} \Psi^{SMm}_{m'}(\vp,-\vq; N). \ee

As seen from Eqs.~\eqref{PsiVec23transp}, \eqref{PsiVecDefParityTransf},
and~\eqref{PsiVec23ParityTransf},
interchange of identical nucleons and space inversion yield
relations between components of
WF~\eqref{PsiVec} with the same quantum numbers $SMmm'$
at different points in the space of Jacobi momenta.

When interaction between nucleons is $\Trev$--even, the WF fulfills
\be\label{PsiVecTime}
\Psi^{SMm}_{m'}(\vp,\vq\,)=
(-1)^{S+M+m-m'}\bigl(\Psi^{S,-M,-m}_{-m'}(-\vp,-\vq\,)\bigr)^*.\ee
For time reversal $\Trev$
we follow the convention $\Trev\bigl|JM\bigr>=(-1)^{J-M}\bigl|J,-M\bigr>,$
where $\bigl|JM\bigr>$ is
an eigenstate of angular momentum operators ${\vec J}^{\;^2}$ and $J_z.$

Time--reversal invariance of nuclear forces yields
for components of the WF with definite parity
\be\label{PsiVecDefParityTime}
\Psi^{SMm}_{-m'}(\vp,\vq; N)=
(-1)^{S+M+m+m'+N}\bigl(\Psi^{S,-M,-m}_{m'}(\vp,\vq; N)\bigr)^*,\ee
where values of WFs are taken at the same point of the $\vp,\vq$--space.
With the help of Eq.~\eqref{PsiVecDefParityTime}
components of tensor~\eqref{PsiVecDefParity} with $m'=-\Oh$
can be obtained from ones having $m'=\Oh.$
Property~\eqref{PsiVecDefParityTime} of the tensor WF was used
in calculations{~\cite{Kotlyar87,KotlyarMelnikShebeko95,Kotlyar05}.}

It is convenient to regard WF~\eqref{PsiVec} as a collection
of four $2\times2$ matrices $\Phi^{SM}(\vp,\vq\,),$ defined as
\be\label{PhiSM}
\left<\Oh m\right| \Phi^{SM}(\vp,\vq\,) \left| \Oh m'\right>=\Psi^{SMm}_{m'}(\vp,\vq\, ),
\qquad (SM=00,1-\!1,10,11). \ee
The matrices $\Phi^{SM}(\vp,\vq\,)$ in the same way as the 3N state
$\left| \Psi;\Oh m'\right>$ split into parts with definite parity
\be\label{PhiDefParityDecomp}
\Phi^{SM}(\vp,\vq\,)=\sum_{N=0,1} \Phi^{SM}(\vp,\vq; N). \ee

Constraints~\eqref{PsiVecDefParityTime} due to space--time inversion symmetries
can be written as
\begin{align}
S&=0: & \text{Im}\:\Spr \Phi(\vp,\vq; 0)&=0, &  \text{Re}\:\Spr \sigma_i \Phi(\vp,\vq; 0)&=0,
\label{SpPhiS0ParityEvenTime}\\
S&=1: & \text{Re}\:\Spr \Phi_i(\vp,\vq; 0)&=0, &  \text{Im}\:\Spr \sigma_i \Phi_k(\vp,\vq; 0)&=0
\label{SpPhiS1ParityEvenTime}\end{align}
for parity--even components and
\begin{align}
S&=0: & \text{Re}\:\Spr \Phi(\vp,\vq; 1)&=0, &  \text{Im}\:\Spr \sigma_i \Phi(\vp,\vq; 1)&=0,
\label{SpPhiS0ParityOddTimeEven}\\
S&=1: & \text{Im}\:\Spr \Phi_i(\vp,\vq; 1)&=0, &  \text{Re}\:\Spr \sigma_i \Phi_k(\vp,\vq; 1)&=0,
\label{SpPhiS1ParityOddTimeEven}\end{align}
for parity--odd ones,
where  $\sigma_i$ are Pauli matrices and $i,k=x,y,z.$
We use the notations $\Phi(\vp,\vq; N)$ for $\Phi^{S=M=0}(\vp,\vq; N),$
and $\Phi_i(\vp,\vq; N)$ for cartesian components of three--vector $\vec{\Phi}(\vp,\vq; N),$
whose contravariant cyclic components are $\Phi^{S=1,M}(\vp,\vq; N).$

Eqs.~\eqref{SpPhiS0ParityEvenTime}--\eqref{SpPhiS1ParityOddTimeEven}
lead to the suggestion on how to gain constraints on WF~\eqref{PsiVec}
when nuclear forces are time--reversal invariant but may violate space parity.
With the purpose to derive the corresponding relations,
that, in contrast to Eq.~\eqref{PsiVecTime},
include WF for one set of Jacobi momenta,
we construct various independent scalar functions
$\phi_{\nu}=\phi_{\nu}(p,q,\xi)$ from components of the WF in the TR
using tensors $\delta_{ik}, \epsilon_{ikl}$ and vectors $\vph,\vqh,\vec\sigma.$
The functions are labeled by $\nu=1,2,\dots.$
We denote $\xi=\vpsvq,$
where $\vph$ and $\vqh$ stand for unit vectors
pointing in the directions of $\vp$ and $\vq.$

Functions $\phi_{\nu}$ can be obtained by contracting matrices~\eqref{PhiSM}
with $W_\nu$ and $\vec{W}_\nu$ given in Table~\ref{Wnu1to16}.
The expressions for matrices  $W_\nu$ and $\vec{W}_\nu$ contain
axial vector $\vec v=\bigl[\vph\times\vqh\;\bigr].$

Four scalar functions can be built up from the WF components with spin $S=0$
\be\label{phiS0} \phi_\nu=\Oh\Spr W_\nu\Phi,  \qquad  (\nu=1,\dots,4). \ee
Components of the WF that have $S=1$ can be used to produce another 12 functions
\be\label{phiS1} \phi_\nu = \Oh\Spr \vec W_\nu\cdot\vec\Phi, \qquad (\nu=5,\dots,16). \ee

Functions~\eqref{phiS0} and \eqref{phiS1} are real
provided that interaction between nucleons is time--reversal invariant.
Requirements
\begin{align}
S&=0: & \text{Im}\: \Spr W_\nu \Phi&=0, & \nu&=1,\dots,4, \label{SpPhiS0WTimeRev}\\
S&=1: & \text{Im}\: \Spr \vec W_\nu\cdot\vec\Phi&=0, &
\nu&=5,\dots,16, \label{SpPhiS1WTimeRev}
\end{align}
impose 16 real constraints on 16 complex components of the tensor
$\Psi^{SMm}_{m'}(\vp,\vq\,).$
Components of WF~\eqref{PsiVec} with $S=0$ and $S=1$
depend on four and 12 real functions~$\psi_\nu=\psi_\nu(p,q,\xi).$
Thus a set of scalar functions $\psi_\nu$ with $\nu=1,\dots,16$
can be used to construct a WF of 3N bound state with $I=\Oh$
for parity violating nuclear forces that respect time-reversal invariance.

\renewcommand{\arraystretch}{1.5}

\begin{table}[h]
\begin{center}
\caption[Table]{\label{Wnu1to16}
Matrices $W_\nu$ and $\vec{W}_\nu$ used together with WF $\Psi^{SMm}_{m'}(\vp,\vq\:)$
for $S=0$ and $1$ in~\eqref{phiS0} and \eqref{phiS1}
to construct the scalar function $\phi_\nu.$
Components $\Psi^{SMm}_{m'}(\vp,\vq; N)$ of the WF
with definite parity $P_\nu=(-1)^N$ contribute into~\eqref{phiS0} and \eqref{phiS1}
}
\begin{tabular}[t]{|ccr|}                       \hline
$\nu$ & $W_\nu$                    & $P_\nu$ \\ \hline
  1   & $I$                        &  1      \\
  2   & $\vec\sigma\cdot\vph     $ & -1      \\
  3   & $\vec\sigma\cdot\vqh     $ & -1      \\
  4   & $-i\vec\sigma\cdot\vec{v}$ &  1      \\ \hline
\end{tabular}
\hspace*{2mm}
\begin{tabular}[t]{|ccr|}                       \hline
$\nu$ & $\vec{W}_\nu$              & $P_\nu$ \\ \hline
  5   & $\vph$        & -1      \\
  6   & $\vqh$        & -1      \\
  7   & $-i\vec v$    &  1      \\
  8   & $\vec\sigma$  &  1      \\
  9   & $\vec\sigma\cdot\vph\;\vph$      &  1      \\
 10   & $\vec\sigma\cdot\vqh\;\vqh     $ &  1      \\  \hline
\end{tabular}
\hspace*{2mm}
\begin{tabular}[t]{|ccr|}                         \hline
$\nu$ & $\vec{W}_\nu$              & $P_\nu$ \\   \hline
 11   & $\frac{1}{2}(\vec\sigma\cdot\vph \; \vqh + \vec\sigma\cdot\vqh \; \vph\:) $
      &  1      \\
 12   & $[\vec\sigma\times\vec{v}\:]$  &  1      \\
 13   & $-\frac{i}{2}(\vec\sigma\cdot\vph \; \vec{v}+\vec\sigma\cdot\vec{v} \; \vph\:) $
      & -1      \\
 14   & $-\frac{i}{2}(\vec\sigma\cdot\vqh \; \vec{v}+\vec\sigma\cdot\vec{v} \; \vqh\:) $
      & -1      \\
 15   & $-\frac{i}{2}[\vec\sigma\times\vph\:]$ & -1      \\
 16   & $-\frac{i}{2}[\vec\sigma\times\vqh\:]$ & -1      \\ \hline
\end{tabular}
\end{center}
\end{table}

Set of matrices $W_\nu$ and $\vec{W}_\nu$ is not unique.
Instead of $\phi_{15}$ and $\phi_{16}$ one can use, e.g.
\be\label{phi1718}
\phi_{17}=\frac{1}{4i}[\vec v\times\vph\:]\cdot\Spr[\vec\sigma\times\vec\Phi],
\qquad\text{and}\qquad
\phi_{18}=\frac{1}{4i}[\vec v\times\vqh\:]\cdot\Spr[\vec\sigma\times\vec\Phi]. \ee
Since $\phi_\nu$ with $\nu=15,\dots,18$ obey
\be\label{phi15161718}
(1-\xi^2)\phi_{15}=-\xi\phi_{17}+\phi_{18} \qquad\text{and}\qquad
(1-\xi^2)\phi_{16}= -\phi_{17}+\xi\phi_{18}, \ee
functions $\phi_{17},$ $\phi_{18}$ generate no new constraints on the tensor WF.

The r.h.s. of \eqref{phiS1} with $\nu=8,\dots,16$ can be written as
$\Oh\Spr W_{ik}\sigma_i\Phi_k,$ where $i,k=x,y,z.$ We use the tensor $W_{ik}$
in symmetrical (antisymmetrical) form for $\nu=8,\dots,11,13,14$ $(\nu=12,15,\dots,18).$
Consideration of diverse constructions for matrices $W_\nu,\vec W_\nu$ or for
tensors $ W_{ik}$ and respective functions $\phi_\nu$ with $\nu>16$
does not extend constraint set~\eqref{SpPhiS0WTimeRev} and~\eqref{SpPhiS1WTimeRev}.
Additional functions, for which $\nu>16,$
and $\phi_\nu$ with $\nu=1,\dots,16$
proved to be not independent.
For example, for $W_{ik}=v_i v_k$ and correspondingly
$\phi_{19}=\Oh\Spr\,\vec{v}\cdot\vec\sigma\;\vec{v}\cdot\vec\Phi$
we have
\be\label{phi19phi}
\phi_{19}=(1-\xi^2)\phi_8-\phi_9-\phi_{10}+2\xi\phi_{11}. \ee
Restrictions on the WF in TR due to time--reversal invariance are exhausted
by equations that are generated with the help of the matrices listed in Table~\ref{Wnu1to16}.


Matrices $W_\nu$ and $\vec{W}_\nu$ presented in Table~\ref{Wnu1to16}
are chosen to have definite parity.
Respectively, functions~\eqref{phiS0},\eqref{phiS1} and
constraints~\eqref{SpPhiS0WTimeRev} and~\eqref{SpPhiS1WTimeRev}
involve even or odd components~\eqref{PsiVecDefParity} of the WF.

Positive--parity functions $\Psi^{SMm}_{m'}(\vp,\vq; 0)$ with $S=0$ and 1
satisfy~\eqref{SpPhiS0WTimeRev}  
and~\eqref{SpPhiS1WTimeRev}.     
Moreover, these functions obey
\begin{align}
S&=0: & \text{Re}\:\Spr W_\nu \Phi&=0, & (\nu&=2,3),            \label{SpPhiS0WTimeRevPeven}\\
S&=1: & \text{Re}\: \Spr \vec{W}_\nu \cdot\vec{\Phi}&=0,
                                       & (\nu&=5,6,13,\dots,16).\label{SpPhiS1WTimeRevPeven}
\end{align}
Eqs.~\eqref{SpPhiS0WTimeRev},\eqref{SpPhiS1WTimeRev},\eqref{SpPhiS0WTimeRevPeven}
and~\eqref{SpPhiS1WTimeRevPeven} place 24 real conditions
on 16 complex components of WF~\eqref{PsiVecDefParity}.
So, even WF $\Psi^{SMm}_{m'}(\vp,\vq; 0)$
depends on eight real functions $\psi_\lambda,$
for which we assign $\lambda=1,\dots,8.$
This observation conforms to conclusions~\cite{Fachruddin04}.

Negative--parity functions $\Psi^{SMm}_{m'}(\vp,\vq;1)$ like $\Psi^{SMm}_{m'}(\vp,\vq;0)$
meet Eqs.~\eqref{SpPhiS0WTimeRev} and~\eqref{SpPhiS1WTimeRev}.
The odd WF components also satisfy~\eqref{SpPhiS0WTimeRevPeven} with $\nu=1,4$
and~\eqref{SpPhiS1WTimeRevPeven} with $\nu=7,\dots,12.$
Taking into account the constraints on parity--odd components, we see that
these components can be build up using eight real functions $\psi_\lambda,$
which we label by  $\lambda=9,\dots,16.$

Eqs.~\eqref{PsiVecDefParityTime}
or~\eqref{SpPhiS0ParityEvenTime}--\eqref{SpPhiS1ParityOddTimeEven}
provide an incomplete set of constraints caused by space--time inversion.
Really, restrictions on even components $\Psi^{SMm}_{m'}(\vp,\vq;0)$ given
neither by~\eqref{SpPhiS0WTimeRev} with $\nu=2,3$
    nor by~\eqref{SpPhiS1WTimeRev} with $\nu=5,6,13,\dots,16$
stem from~\eqref{PsiVecDefParityTime}.
Conditions~\eqref{SpPhiS0WTimeRev} with $\nu=1,4$
       and~\eqref{SpPhiS1WTimeRev} with $\nu=7,8,\dots,12$
for odd components $\Psi^{SMm}_{m'}(\vp,\vq;1)$
cannot be deduced form~\eqref{PsiVecDefParityTime} either.


Set of constraints on components of the tensor WF,
which is placed by requirement of time-reversal invariance,
degenerates, and number of independent relations between the WF components decreases,
when relative momenta are collinear, i.e. $\vph=\pm p\vqh.$
Constraints and structure of the WF for this case are examined in \ref{AppCollinearRelMom}


\section{Decomposition of the Wave Function over Polarization Operators}

Aiming to elucidate structure of WF~\eqref{PsiVec},
which is a reducible tensor, we decompose it
over the complete system of polarization operators (POs) $T^{KM}(s)$ for spin $s=\Oh.$

Properties of the POs are discussed, e.g. in~\cite{Varshalovich,Ohlsen,Nemets}.
We take POs in the conventions of Ref.~\cite{Varshalovich}.
The contravariant POs are given by
\be\label{POContVCov} T^{KM}(s)=(-1)^MT_{K,-M}(s). \ee
The matrices of the covariant POs are
\be\label{POmatr}
\left<sm\right| T_{KM}(s) \left|sm'\right>=(2K+1)^{\Oh}(2s+1)^{-\Oh}\;C^{sm}_{sm'\;KM}.\ee

In terms of matrices~\eqref{PhiSM} the decomposition of tensor WF~\eqref{PsiVec} reads
\be\label{PhiSMDecomp}
\Phi^{SM}(\vp,\vq\,)=\sqrt2\sum_{K=0,1;M}
\bar{\Psi}^{SM}_{\hphantom{SM}KM'}(\vp,\vq\,) \; T^{KM'}(\Oh). \ee

Tensor $\bar{\Psi}^{SM}_{\hphantom{SM}KM'}(\vp,\vq\,)$
is to be constructed from the Jacobi momenta $\vp$ and $\vq.$
Component of the tensor with $S=K=0$ is a scalar function
\be\label{PsiS0K0}
\bar{\Psi}^{00}_{\hphantom{00}00}(\vp,\vq\,)=\psi_1(p,q,\xi).\ee
As above, we denote $\psi_\lambda=\psi_\lambda(p,q,\xi),$ where $\lambda=1,2,\dots.$
The tensors with $S=0,\;K=1$ or $S=1\;K=0,$  which are equivalent to three--vectors,
can be written as
\begin{align}
\bar{\Psi}^{00}_{\hphantom{00}1M}(\vp,\vq\,)&=
\sqrt2 \; v'_M \; \psi_2
+\hat{p}_M \; \psi_9+\hat{q}_M \; \psi_{10},            \label{PsiS0K1} \\
\bar{\Psi}^{1M}_{\hphantom{1M}00}(\vp,\vq\,)&=
\sqrt2 \; v'^M \; \psi_3
+\hat{p}^M \; \psi_{11}+\hat{q}^M \; \psi_{12},         \label{PsiS1K0} \end{align}
where pseudovector $v'_M=\bigl\{\vph\otimes\vqh\bigr\}_{1M}. $

Irreducible tensor product that has rank $J$
of two irreducible tensors $A_{J'M'}$ and $B_{J''M''}$ is~\cite{Varshalovich}
\be\label{TensProd}
\bigl\{A_{J'} \otimes B_{J''} \bigr\}_{JM}=\sum_{M'M''}\;
C^{JM}_{J'M'\;J''M'} \; A_{J'M'}\;B_{J''M''}. \ee
The contravariant and covariant components of~\eqref{TensProd}
are related by
(cf.\ Eq.~\eqref{POContVCov})
\be\label{TensProdContVCov}
\bigl\{A^{J'} \otimes B^{J''} \bigr\}^{JM}=
(-1)^M\;\bigl\{A_{J'} \otimes B_{J''} \bigr\}_{J,-M}. \ee

Structure of the reducible tensor $\bar{\Psi}^{1M}_{\hphantom{1M}1M'}(\vp,\vq\,)$
can be determined from decomposition
\be\label{PsiS1K1}
\bar{\Psi}^{1M}_{\hphantom{1M}1M'}(\vp,\vq\,)=
\sum_{k=0,1,2;\;\kappa} \; w_{k\kappa}(\vp,\vq\,) \;
\bigl<1M\bigr|T^{k\kappa}(1)\bigl|1M'\bigr>, \ee
where, similarly to Eqs.~\eqref{PsiS0K0}--\eqref{PsiS1K0},  one has
\begin{align}
w_{00}(\vp,\vq\,)\;&= \sqrt3 \; \psi'_4,                                    \label{wk00} \\
w_{1M}(\vp,\vq\,)&= 2 v'_M \; \psi'_5
- \sqrt2 \hat{p}_M \; \psi_{13} - \sqrt2 \hat{q}_M \; \psi_{14}. \label{wk1M} \end{align}

Term $w_{2\kappa}(\vp,\vq\,)$ in~\eqref{PsiS1K1}
receives contributions from irreducible tensor products of rank two,
which may be quadratic and cubic in relative momenta,
\be\label{wk2kappa}\begin{split}
w_{2\kappa}(\vp,\vq\,)&=
- \bigl\{\vph\otimes\vph\bigr\}_{2\kappa} \; \psi_6
- \bigl\{\vqh\otimes\vqh\bigr\}_{2\kappa} \; \psi_7
-2\bigl\{\vph\otimes\vqh\bigr\}_{2\kappa} \; \psi'_8-          \\ &
-2\sqrt2 \bigl\{\vph\otimes v'\bigr\}_{2\kappa} \; \psi_{15}
-2\sqrt2 \bigl\{\vqh\otimes v'\bigr\}_{2\kappa} \; \psi_{16}. \end{split}\ee

Eqs.~\eqref{PsiS0K0}--\eqref{PsiS1K0}, \eqref{PsiS1K1}--\eqref{wk2kappa}
yield for components of the WF that have positive parity
\begin{align}
\Phi(\vp,\vq; N=0)&=\psi_1 + i\vec{\sigma}\cdot\vec{v} \; \psi_2,
\label{PhiN0}\\
\vec{\Phi}(\vp,\vq; N=0)&=i\vec{v} \; \psi_3 + \vec{\sigma} \; \psi_4
+\vph \; \vec{\sigma}\cdot\vqh \; \psi_5 +\vph \; \vec{\sigma}\cdot\vph \; \psi_6 +
\nonumber\\&
+\vqh \; \vec{\sigma}\cdot\vqh \; \psi_7 +\vqh \; \vec{\sigma}\cdot\vph \; \psi_8.
\label{PhiVecN0} \end{align}

Matrices $\Phi^{SM}(\vp,\vq; N=0)$
depend on functions $\psi_\lambda$ with $\lambda=1,\dots,8,$ where
\be\label{psipsiprime}
\psi_4=\psi'_4-{\tss{\frac{1}{3}}} (\psi_6+\psi_7+2\xi\psi'_8), \qquad
\psi_5= \psi'_5+\psi'_8,                                    \qquad
\psi_8=-\psi'_5+\psi'_8. \ee

Other 8 functions $\psi_\lambda$ with $\lambda=9,\dots,16$
appear in the negative parity contributions to the WF.
Component $\Psi^{SMm}_{m'}(\vp,\vq; N=1)$ with zero spin $S$ reads
\be\label{PhiN1}
\Phi(\vp,\vq; N=1)=\vec{\sigma}\cdot\vph\;\psi_9 + \vec{\sigma}\cdot\vqh\;\psi_{10}. \ee
The part of the WF with $S=1$ can be written as
\be\label{PhiVecN1}\begin{split}
\vec{\Phi}(\vp,\vq; N=1)&=\vph\;\psi_{11}+\vqh\;\psi_{12}
+i\bigl[\vec{\sigma}\times\vph\:\bigr]\;\psi_{13}
+i\bigl[\vec{\sigma}\times\vqh\:\bigr]\;\psi_{14}+ \\&
+i\bigl(\vph\;\vec{\sigma}\cdot\vec{v}+\vec{v}\;\vec{\sigma}\cdot\vph \:\bigr)\;\psi_{15}
+i\bigl(\vqh\;\vec{\sigma}\cdot\vec{v}+\vec{v}\;\vec{\sigma}\cdot\vqh \:\bigr)\;\psi_{16}.
\end{split}\ee

Eqs.~\eqref{PhiN0},\eqref{PhiVecN0},\eqref{PhiN1}, and \eqref{PhiVecN1} express
WF $\Psi^{SMm}_{m'}(\vp,\vq\,)$ of 3N bound state
with total angular momentum $I=\Oh$ in terms of
16 scalar functions $\psi_\lambda(p,q,\xi),$ which are real
for $\Trev$--even nuclear forces.
As show in Appendix~B,
functions $\psi_\lambda$ are related to $\phi_\lambda,$ where $\lambda=1,\dots,16,$
by linear transformation. The case of collinear vectors $\vp$ and $\vq,$
when the transformation has degenerated matrix, is discussed in  Appendix~A.


\section{Operator Form of the Wave Function}

Relation between TR and OFs for the WF can be obtained from decomposition
\be\label{WFVecDecomp01}
\left|\Psi;\Oh m'\right>=\sum_{SMm} \; \int \; d^3p \; d^3q \;
\Psi^{SMm}_{m'}(\vp,\vq\,) \; {{\left|\vp\:\vq; SM\Oh m\right>}{_{23,1}}}. \ee
R.h.s. of~\eqref{WFVecDecomp01} contains
spin states ${\left|SM\right>}{_{23}}$ with $S\!\!=\!\!0$ and 1, which are related~by
\be\label{S1Msigma23S00} \left|1M\right>_{23}=\sigma_M(23)\left|00\right>_{23}, \ee
where operator              $ \vec\sigma(23)=\Oh(\vec\sigma(2)-\vec\sigma(3))  $
was introduced in~\cite{GerjuoySchwinger}.
The covariant cyclic components~\cite{Varshalovich} of the vector $\vec\sigma(23)$
are denoted by $\sigma_M(23).$
With the help of identity~\eqref{S1Msigma23S00}
one can cast \eqref{WFVecDecomp01} into the form
\be\label{WFVecDecomp0}
\left|\Psi;\Oh m'\right>= \int \; d^3p \; d^3q \;
{\hat\Psi}(\vp,\vq\,) \; {{\left|\vp\:\vq; \; S=M=0,\Oh m\right>}{_{23,1}}}. \ee

Operator ${\hat\Psi}(\vp,\vq\,)$ acting in the spin space
can be expressed through the matrices $\Phi^{SM}(\vp,\vq;N),$ viz.,
\be\label{OpPsi0Phi}
{\hat\Psi}(\vp,\vq\,)=\sum_{N=0,1} \;
\bigl( \Phi(\vp,\vq;N) + \vec{\Phi}(\vp,\vq;N)\cdot\vec\sigma(23) \bigr). \ee
In Eq.~\eqref{OpPsi0Phi} the matrices $\Phi^{SM}(\vp,\vq;N)$
are given by expressions~\eqref{PhiN0}, \eqref{PhiVecN0}, \eqref{PhiN1}, \eqref{PhiVecN1}
where $\vec{\sigma}$ is substituted by $\vec{\sigma}(1).$

Representation~\eqref{OpPsi0Phi} can be also derived
by means of less formal decomposition of ${\hat\Psi}(\vp,\vq\,)$
in terms of the cartesian components of the operators $\vec\sigma(1)$ and $\vec\sigma(23)$
\be\label{hatPsiSpinOpvpvq}\begin{split}
\hat\Psi(\vp, \vq\:)&=\Psi(\vp,\vq\:;1)
+\sum_n \bigl( \Psi_n(\vp,\vq\:;2) \; \sigma_n(1)
+              \Psi_n(\vp,\vq\:;3) \; \sigma_n(23) \bigr) +
\\ &
+\sum_{nn'} \Psi_{nn'}(\vp,\vq\:) \; \sigma_n(1)\sigma_{n'}(23), \qquad (n,n'=x,y,z).
\end{split}\ee
Scalar $\Psi(\vp,\vq\:;\lambda=1),$ vectors $\Psi_n(\vp,\vq\:;\lambda=2,3),$
and tensor $\Psi_{nn'}(\vp,\vq\:)$
are to be build up in terms of the vectors $\vph,\vqh$
and scalar functions $\psi_\lambda(p,q,\xi).$

In analogy to~\eqref{PsiS0K0}--\eqref{PsiS1K0} the scalar and the vectors
can be written as
\be\label{Psi123psi1291031112}\begin{split}
\Psi(\vp,\vq\:;1)=\psi_1,  \qquad\qquad            
&
\vec{\Psi}(\vp,\vq\:;2)
=i\vec{v}\;\psi_2+\vph\;\psi_9   +\vqh\;\psi_{10}, 
\\&
\vec{\Psi}(\vp,\vq\:;3)
=i\vec{v}\;\psi_3+\vph\;\psi_{11}+\vqh\;\psi_{12}. 
\end{split}\ee
The second rank tensor can be split into symmetrical and antisymmetrical parts
\be\label{Psinn}
\Psi_{nn'}(\vp,\vq\:)=\delta_{nn'}\;\psi_4 +\Psi^s_{nn'}+\Psi^a_{nn'}. \ee

We may take the symmetrical tensor in the form
\be\label{PsinnSym}\begin{split}
\Psi^s_{nn'}(\vp,\vq\:)&=
\hat{p}_n\hat{p}_{n'}\;\psi_6+\hat{q}_n\hat{q}_{n'}\;\psi_7+
(\hat{p}_n\hat{q}_{n'}+\hat{q}_n\hat{p}_{n'})\;\psi'_8+
\\&+i\;(v_{n}\hat{p}_{n'}+\hat{p}_{n}v_{n'})\psi_{15}+
    i\;(v_{n}\hat{q}_{n'}+\hat{q}_{n}v_{n'})\psi_{16}.
\end{split}\ee
Inclusion of terms that contain more than three unit vectors into~\eqref{PsinnSym}
leads to redefinition of functions $\psi_\nu,$
e.g., adding tensor $v_n v_{n'}$ modifies
function $\psi_\nu$ with $\nu=4,6,7$ and $\psi'_8.$

The antisymmetrical tensor in~\eqref{Psinn} can be given by
\be\label{PsinnAntiSym}
\Psi^a_{nn'}(\vp,\vq\:)=
-\sum_{l=x,y,z}\epsilon_{nn'l}\;(v_l\;\psi_5+ip_l\;\psi_{13}+iq_l\;\psi_{14}). \ee
Insertion of other terms, for example
$ i\epsilon_{nn'l} \; \epsilon_{lkk'} \; v_k \; p_{k'} $
or
$ i\epsilon_{nn'l} \; \epsilon_{lkk'} \; v_k \; q_{k'},$
into~\eqref{PsinnAntiSym} changes $\psi_\nu$ with $\nu=13,14$
leaving structure of the r.h.s. unaltered.

Operator $\hat\Psi(\vp, \vq\:)$
being defined in Eqs.~\eqref{hatPsiSpinOpvpvq}--\eqref{PsinnAntiSym}
conforms to expressions~\eqref{PhiN0}, \eqref{PhiVecN0}, \eqref{PhiN1},
\eqref{PhiVecN1}, and~\eqref{OpPsi0Phi} for matrices $\Phi^{SM}(\vp,\vq;N).$

Representations of the WF in terms of the POs as well as with the use of cartesian tensors
allow one to relate functions $\psi_\lambda(p,q,\xi)$ and
components of the WF with definite values of total orbital angular momenta.
As seen from Eqs.~\eqref{PhiN0} and \eqref{PhiVecN0}, the $S$--state $^2S_\oh$ originates
not merely from $\psi_{\lambda=1}$ and $\psi_{\lambda=4}.$
As far as in Eqs.~\eqref{PhiVecN0} or \eqref{Psinn}, \eqref{PsinnSym}
symmetrical traceless tensor
\be\label{PsinnSymTr0}
\Psi^{s,0}_{nn'}(\vp,\vq\:)=\Psi^{s}_{nn'}(\vp,\vq\:)
-\ott\delta_{nn'}\;(\psi_6+\psi_7+2\xi\psi'_8) \ee
is not singled out,
$\psi_{\lambda=6,7}$ and $\psi'_{\lambda=8}$ contribute to the $S$--state,
and through tensor $\Psi^{s,0}_{nn'}(\vp,\vq\:)$ these functions are related
to $D$--wave  $^4D_\oh.$
Pseudovector $\vec v$ in~\eqref{PhiN0} and~\eqref{PhiVecN0}
together with antisymmetrical part of~\eqref{PhiVecN0} determine $P$--waves.
Component $^2P_\oh$ depends on $\psi_{\lambda=2},$
while  $^4P_\oh$ state is due to both $\psi_{\lambda=3},$ and $\psi'_{\lambda=5}.$

Odd components $\Phi^{SM}(\vp,\vq;N=1)$ with spin values $S=0$ and $1,$
which contain $\psi_\lambda$ with $\lambda=9,10$ and $\lambda=11,\dots,14,$
produce negative--parity $^2P_\oh$ and $^4P_\oh$ states, respectively.

The irreducible pseudotensors
$\bigl\{\vph\otimes\bigl\{\vph\otimes\vqh\bigr\}_1\bigr\}_{2\kappa}$ and
$\bigl\{\vqh\otimes\bigl\{\vph\otimes\vqh\bigr\}_1\bigr\}_{2\kappa},$
which have rank two and involve three relative momenta,
together with functions $\psi_{\lambda=15,16}$ (see Eq.~\eqref{wk2kappa})
generate $\Pinv$--odd  $^4D_\oh$ components of the WF.

Eqs.~\eqref{OpPsi0Phi} and~\eqref{hatPsiSpinOpvpvq} can be written as
\be\label{OpPsi0ul}
{\hat\Psi}(\vp,\vq\,)=\sum_{\lambda=0,\dots,16}
 \; \psi_\lambda(p,q,\xi) \; u_\lambda\bigl(\vph,\vqh\,\bigr), \ee
where the spin--angular operators $u_\lambda\bigl(\vph,\vqh\,\bigr)$
corresponding to the scalar functions are separated out.
The operators and functions in~\eqref{OpPsi0ul} are superpositions
of ones in Gerjuoy--Schwinger representation~\cite{GerjuoySchwinger,Fachruddin04}.

Another way to get an OF for the WF is to transform decomposition~\eqref{WFVecDecomp01}
eliminating explicit contribution of the spin state $\chi_0=\left|S=M=0\right>_{23}.$
Identity
\be\label{sigma23chiM} \sigma^M(23) \; \chi_{M'} = \delta_{MM'} \; \chi_0, \ee
where $\chi_M=\left|S=1,M\right>_{23},$ can be used with this end.
Instead of Eq.~\eqref{WFVecDecomp0} one has
\be\label{WFVecDecomp1}
\left|\Psi;\Oh m'\right>= \int \; d^3p \; d^3q \; \left|\vp\:\vq\,\right> \;
\vec{X}(\vp,\vq\,) \cdot \vec\chi \; \chi_{\oh m'}, \ee
where covariant cyclic components of $\vec\chi$ are $\chi_M,$
and $\chi_{\oh m}=\left|\Oh m\right>_1.$
Expression for the operator $\vec{X}(\vp,\vq\,)$ generating 3N--state~\eqref{WFVecDecomp1}
involves $2\times2$ matrix $\vec{\Gamma}(\vp,\vq; N),$
which comes from the term $\Psi^{S=M=0.m}_{m'}(\vp,\vq\,) \; \chi_0$
in~\eqref{WFVecDecomp01},
\be\label{vecX}
\vec{X}(\vp,\vq\,)=\sum_{N=0,1} \;
\bigl(\vec{\Gamma}(\vp,\vq; N)+\vec{\Phi}(\vp,\vq; N) \bigr).\ee

Parity-even component of $\vec\Gamma$ is given by
\be\label{GammaN0}\begin{split}
\vec{\Gamma}(\vp,\vq; N=0)&=(3a_1+a_2+a_3)^{-1} \times
\\&
\times \bigl( a_1 \; \vec{\sigma}(23)
+ a_2 \; \vec{\sigma}(23)\cdot\vph \; \vph
+ a_3 \; \vec{\sigma}(23)\cdot\vqh \; \vqh \bigr) \; \psi_1 +
\\&
+i/2 \bigl( \vec{\sigma}(1)\cdot\vec{\sigma}(23) \; \vec{v}
+\vec{\sigma}(23)\cdot\vec{v} \; \vec{\sigma}(1) \bigr)  \; \psi_2. \end{split}\ee

Representation~\eqref{GammaN0} for $\vec{\Gamma}(\vp,\vq; N=0)$ is not unique.
Choice of coefficients $a_i$ $(i=1,2,3)$ depends on the purpose of the transformation.

Parity--odd $^2P_\oh$ state
\be\label{GammaN1}\begin{split}
\vec{\Gamma}(\vp,\vq; N=1)&=
\bigl( \vec{\sigma}(1) \cdot\vec{\sigma}(23) \; \vph
      +\vec{\sigma}(23)\cdot\vph             \;  \vec{\sigma}(1)
\bigr) \; \psi_9/2 +                                                 \\&+
\bigl( \vec{\sigma}(1) \cdot\vec{\sigma}(23) \; \vqh
      +\vec{\sigma}(23)\cdot\vqh             \;  \vec{\sigma}(1)
\bigr) \; \psi_{10}/2                                                \end{split}\ee
originates from~\eqref{PhiN1} and contains no ambiguities.


\section{Common Features of Tensor and Operator Representations \\
for Deuteron and \He Wave Functions}

In calculations of reaction amplitudes, e.g. of \He two--body photodisintegration, or
the momentum distributions of proton--deuteron clusters in \He
both WFs of 2N and 3N nuclei are constructed making use of partial--wave decompositions.
The deuteron WF is known~\cite{Arenhoevel,Liu035501,Liu045501,Schiavilla}
to contain ${^{2S+1}L}_J={^3S}_1,$ ${^1P}_1,$ ${^3P}_1,$ ${^3D}_1$ states
when nucleon--nucleon potential includes both $\Pinv$--even and $\Pinv$--odd contributions.
In the case of a three--body bound state partial--wave series
involve infinite number of terms~\cite{Glockle83}.

WFs of deuteron and \He in the TR consist of 9 and 16 complex components
regardless of whether nuclear forces conserve parity.
The components of WFs in the TR are not independent.

In the center of mass system WF of deuteron with the total angular momentum $J=1$
and its projection $M'$ reads
\be\label{DetrVec}
\phi^{SM}_{\hphantom{SM}1M'}(\vp\,)={_{23}\!\left<\vphantom{\phi;JM'}
\vp; SM\right|}\left.\phi;J=1,\,M'\right>, \qquad (S=0,1).\ee
Nucleons in deuteron have number labels 2 and 3.

Component of the WF~\eqref{DetrVec} with $S=0$ can be written as
\be\label{DetrVecS0}
\phi^{S=M=0}_{\hphantom{SM}1M'}(\vp\,)=\hat{p}_{M'} \; \phi({^1P}_1), \ee
where notation $\phi({^{2S+1}L}_J)=\phi(p;{^{2S+1}L}_J)$ is used.

Tensor $\phi^{S=1,M}_{\hphantom{SM}1M'}(\vp\,)$ is a $3\times3$ matrix,
which can be decomposed similarly to $\bar{\Psi}^{SM}_{\hphantom{SM}KM'}(\vp,\vq\,)$
over the POs with spin $s=1$ (see Eq.~\eqref{PsiS1K1}).
Coefficients in the decomposition,
which are denoted by $\phi_{k\kappa}(\vp\,),$ where $k=0,1,2,$
can be constructed as follows
\be\label{DetrVecS1}\begin{split}
\phi_{00}&=\sqrt3 \; \phi'({^3S}_1),                          \qquad\qquad\qquad
\phi_{1\kappa}=\sqrt2 \; \hat{p}_\kappa \; \phi({^3P}_1),
\\
\phi_{2\kappa}&=-\bigl\{\vph\otimes\vph\bigr\}_{2\kappa} \; \phi({^3D}_1).
\end{split}\ee
Time--reversal invariance of NN interaction implies that
the functions $\phi(p;{^{2S+1}L}_J)$ are real.

Part of the deuteron WF with $S=1$ is
\be\label{DetrVecS1AB}\begin{split}
\sum_{MM'} & A_M B^{M'} \; \phi^{1M}_{\hphantom{1M}1M'}(\vp\,)= \\
&=\vec{A}\cdot\vec{B} \; \phi({^3S}_1)
+i \;\vph\cdot \bigl[ \vec{A}\times\vec{B} \bigr] \; \phi({^3P}_1)
+\vec{A}\cdot\vph \;\; \vec{B}\cdot\vph \; \phi({^3D}_1),    \end{split}\ee
where $\phi({^3S}_1)=\phi'({^3S}_1)-\phi({^3D}_1)/3,$
and auxiliary vectors $\vec{A}$ and $\vec{B}$ are introduced for convenience.

Eq.~\eqref{DetrVecS1AB} can be written in terms of cartesian components
\be\label{DeutTensParity} \phi_{nn'}(\vp\,)=\sum_{N=0,1} \; \phi_{nn'}(\vp; N) \ee
of the tensor $ \phi^{1M}_{\hphantom{1M}1M'}(\vp\,).$
The $\Pinv$--even part of~\eqref{DeutTensParity}
\be\label{DeutTensEven}
\phi_{nn'}(\vp; N=0)=\delta_{nn'} \; \phi'({^3S}_1)
+ \hat{p}_n \hat{p}_{n'} \; \phi({^3D}_1) \ee
is symmetrical in indices $n,n'$ and real.
The $\Pinv$--odd contribution to the WF
originating from ${^1P}_1$ wave~\eqref{DetrVecS0} is real,
whereas one coming from the ${^3P}_1$ wave is antisymmetrical and imaginary
\begin{align}
S&=0: & \phi_{n}(\vp\,)&=\hat{p}_n \; \phi({^1P}_1), \label{DeutTens1P1}\\
S&=1: & \phi_{nn'}(\vp; N=1)&=i \; \epsilon_{nn'l} \; \hat{p}_l \; \phi({^3P}_1).
\label{DeutTens3P1}
\end{align}

Eqs.~\eqref{DetrVecS0}, \eqref{DetrVecS1}  and~\eqref{DeutTensEven}--\eqref{DeutTens1P1}
correspond to partial--wave decompositions~\cite{Arenhoevel,Liu035501,Liu045501,Schiavilla}
of WF~\eqref{DetrVec}.
Compact representation~\eqref{DeutTensEven}--\eqref{DeutTens1P1} for the deuteron WF
prompts us to search for corresponding constructions in the case of 3N and 4N systems.

OFs for the deuteron WF can be derived
from expressions~\eqref{DeutTensEven}--\eqref{DeutTens3P1}.
Vector $\bigl| \vec{\phi}\,\bigr>$ with cyclic components that
coincide with the deuteron state $\left|\phi;J=1,\,M\right>$ is produced according to
\be\label{DeutVectOpDec0}
\bigl|\vec{\phi}\,\bigr>=\int \; d^3p \; \vec{\phi}(\vp\,) \;
{{\left|\vp; \; S=M=0\right>}{_{23}}} \ee
by operator $\vec{\phi}(\vp\,)$ acting in the spin space.
Axial-- and polar--vector parts of
\be\label{DeutVectOpPhi0}\begin{split}
\vec{\phi}(\vp\,) &= \vec{\sigma}(23) \; \phi({^3S}_1) + \vph \; \phi({^1P}_1)+ \\&
+i \; \bigl[ \vec{\sigma}(23)\times\vph \, \bigr] \; \phi({^3P}_1)
+\vec{\sigma}(23)\cdot\vph \; \vph \; \phi({^3D}_1)                   \end{split}\ee
spring from parity conserving and violating nucleon--nucleon interaction.
The operator $\vec{\phi}(\vp\,)$ is applied in~\eqref{DeutVectOpDec0}
at the two-nucleon state ${\left|S=M=0\right>}{_{23}}$ with zero total spin.
In this respect, Eq.~\eqref{DeutVectOpDec0}
is similar to decomposition~\eqref{WFVecDecomp0} in the case of 3N nuclei.
While transformations of~\eqref{DeutVectOpDec0} under rotations
are provided by the operator $\vec{\phi}(\vp\,),$
in the case of 3N bound state, vector ${\left|\Oh m'\right>}_1$
is responsible for the correct properties of $\left|\Psi; \Oh m'\right>,$
which is generated by the scalar operator $\hat{\Psi}(\vp,\vq\,).$

As seen from~\eqref{DeutVectOpPhi0} and~\eqref{PhiVecCollin},
both the deuteron WF and the $S=1$ component of 3N WF for collinear relative momenta
depend on four scalar functions and possess the same space--spin structure.

One can get from~\eqref{DetrVecS0},\eqref{DetrVecS1} and~\eqref{DeutVectOpDec0}
other representation
\be\label{DeutVectOpDec1}
\bigl|\vec{\phi}\,\bigr>=\int \; d^3p \; \vec{\chi}(\vp\,) \; \left|\vp\,\right> \ee
that involves explicitly only the 2N spin states $\vec{\chi}$ with S=1.
Vector in the spin space $\vec{\chi}(\vp\,)$ is given by
\be\label{DeutVectOpChi1}\begin{split}
\vec{\chi}(\vp\,)&=\vec{\chi} \; \phi({^3S}_1)
+ \Oh\bigl(\vec{\sigma}(23)\cdot\vph \;\; \vec{\chi}
+     \vec{\sigma}(23) \;\; \vph\cdot\vec{\chi}\,\bigr) \; \phi({^1P}_1) + \\&
+i \; \bigl[ \vec{\chi}\times\vph \, \bigr] \; \phi({^3P}_1)
+\vec{\chi}\cdot\vph \;\; \vph \; \phi({^3D}_1).                   \end{split}\ee

Identity~\eqref{sigma23chiM} has been employed
to obtain contribution of ${^1P}_1$ state to~\eqref{DeutVectOpChi1}.
Derivation of Eqs.~\eqref{DeutVectOpDec1} and~\eqref{DeutVectOpChi1}
does not suffer from any ambiguities in contrast to
the transformation of the $S=0$ contribution to the 3N WF that
yields the matrix $\vec{\Gamma}(\vp,\vq; N=0).$
Really, angular dependence of ${^1P}_1$ component~\eqref{DetrVecS0}
is determined by only one vector $\vph,$ and
application of the identity is straightforward.


\section{Summary and Outlook}

We discuss TR  
       and OFs 
for WF of 3N bound state with the total angular momentum $I=\Oh.$
The WF in TR 
is a reducible tensor $\Psi^{SMm}_{m'}(\vp,\vq\,),$
which consists of 16 complex components.
Time--reversal invariance implies that these components are not independent and
yields 16 real constraints imposed on the tensor WF.
The WF in TR, e.g. of \Hep, can be built up using 16 real scalar functions
$\psi_\lambda(p,q,\vph\cdot\vqh\,).$

Structure of the tensor $\Psi^{SMm}_{m'}(\vp,\vq\,)$ is scrutinized
with the help of decompositions over POs.
Under the assumption that interaction between nucleons is
$\Trev$--even, 
the WF is expressed in terms of 16 functions $\psi_\lambda.$
Eight of them  with $\lambda=1,\dots,8$ produce parity--even components of the WF.
Ones with $\lambda=9,\dots,16$ appear for parity non--conserving nuclear forces and
cause parity--odd admixtures in the WF.
Functions  $\psi_\lambda$ are  related to
the 3N sates with definite values of total angular orbital momenta.

To gain an OF for the 3N bound state we proceed from the WF in TR.
In one of the derived OFs the spin--angular structure of the WF
is generated by operators $u_\lambda\bigl(\vph,\vqh\,\bigr),$ where $\lambda=1,\dots,16,$
applied at the spin state ${\left| S\!=\!M\!=\!0,\,\Oh m\right>}{_{23,1}}$
with zero total spin $S$ in the pair of identical nucleons.
The unit vectors $\vph,\vqh$ are combined with the nucleon spin operators
to form $u_\lambda,$
which are scalars with respect to rotations in space of the Jacobi momenta.
The operators  $u_\lambda$ and functions $\psi_\lambda$
can be written as superpositions of the respective quantities
in the Gerjuoy-Schwinger representation~\cite{GerjuoySchwinger,Fachruddin04} for the WF.
We note that for collinear relative momenta
components of the 3N WF with $S=1$
assume the form which is similar to one inherent to the deuteron WF.

The deuteron WF, which has total angular momentum $J=1,$
can be constructed from the NN spin states ${\left| SM \right>}{_{23}}$
with total spin either $S=0$ or $S=1.$
The OFs for the deuteron WF are equivalent to
partial--wave decompositions~\cite{Arenhoevel,Liu035501,Liu045501,Schiavilla}.
The obtained OFs differ from
Rarita-Schwinger representation~\cite{RaritaSchwinger,Fachruddin01}.

The 3N WF can be also written in the form of
an operator $\vec{X}(\vp,\vq\,)$ applied at the state
${\left| S\!=\!1,M \; \Oh m\right>}{_{23,1}}.$
The operator $\vec{X}$ transforms like a vector under the rotations.
With several ways existing for obtaining contributions from the term
$\psi_{\lambda=1} \; {\left| S\!=\!M\!=\!0,\;\Oh m\right>}{_{23,1}}$
to $^{2\Ss+1}\Ls_I={^2S}_\oh$ component of the WF,
this alternative OF is not unique.

In~\cite{GerjuoySchwinger,Fachruddin04} and in the present paper
the isospin formalism is not employed.
TR~\eqref{PsiVec} turned out~\cite{Kotlyar05} to be convenient
to derive an OF for the WF taking into account isospin degrees of freedom.
Explicitly antisymmetrical operator representation for the WF within the isospin formalism
can be deduced from the decomposition
$\bigl| \Psi\bigr>=(1-(1,2)-(1,3))\bigl| \Psi^{(1)}\bigr>,$
where $\bigl| \Psi^{(1)}\bigr>$ is taken in an OF.
Transpositions of nucleon quantum numbers in momentum, spin and isospin space
are denoted by $(i,j).$
Being represented in one of the above discussed OFs, vector $\bigl| \Psi^{(1)}\bigr>$
meets the requirement
$(2,3)\bigl| \Psi^{(1)}\bigr>=-\bigl| \Psi^{(1)}\bigr>,$
which provides antisymmetrization of the 3N bound state.

Other approach that allows one to arrive at evidently antisymmetrical WF
is widely known~\cite{Verde,Derrick,SitenkoKharchenko71,Blankleider}.
The WF can be built up from functions $\Psi^{[\nu]}(\vp,\vq\,),$
spin, and isospin states that belong to irreducible representations
($\nu=$ symmetrical, antisymmetrical, and mixed) of the symmetric group $S_3.$
The corresponding expressions for the $\Pinv$--even WF
remain unaltered when effects of parity violation are incorporated,
while functions  $\Psi^{[\nu]}(\vp,\vq\,)$ are to be modified
with the aim to include $\Pinv$--odd contributions.

Owing to progress in calculations~\cite{Nogga02,Nogga03,Viviani06}
of WFs for three--nucleon bound states with realistic nuclear forces,
one can hope that
the 3N nuclei provide an important test laboratory for
studying various processes due to neutral or charged weak currents.
Both TR and OFs for the WFs are convenient for
analysing qualitative features of corresponding matrix elements and
could be useful, e.g. in studying
parity--odd spin--dependent momentum distributions of nucleons in polarized 3N nuclei,
anapole moments of the nuclei,
observables in parity--violating electron scattering on \Hep.


\appendix
\setcounter{section}0
\renewcommand{\thesection}{Appendix \Alph{section}.}
\renewcommand{\theequation}{\Alph{section}.\arabic{equation}}

\setcounter{equation}0
\section{Tensor wave function in the case  \\ of collinear relative momenta}
\label{AppCollinearRelMom}

Number of independent constraints imposed on the WF in TR decreases
when vectors of relative momenta are parallel or antiparallel.
Under conditions, when $\vp=\pm p\vqh,$
one polar and one axial vectors $\vqh$ and $\vec\sigma$ are available
to construct matrices $W_\nu$ and $\vec{W}_\nu.$

Below we discuss the case $\vp=p\vqh.$
One can build up 6 independent scalar functions $\phi_\nu(p,q,1)$
using $\vqh,$ $\vec\sigma,$ $\Phi(p\vqh,\vq\:)$ and $\vec\Phi(p\vqh,\vq\:).$
We define $\phi_\nu$ according
     to~\eqref{phiS0} with $\nu=1,2$
and via~\eqref{phiS1} with $\nu=5,8,9,15.$
Matrices $W_\nu$ and $\vec{W}_\nu$ in~\eqref{phiS0} and~\eqref{phiS1}
are taken from Table~\ref{Wnu1to16}.
Time--reversal invariance yields conditions~\eqref{SpPhiS0WTimeRev} with $\nu=1,2$
                                        and~\eqref{SpPhiS1WTimeRev} with $\nu=5,8,9,15$
that are to be placed on WF $\Psi^{SMm}_{m'}(p\vqh,\vq\, ).$

Tensor WF $\Psi^{SMm}_{m'}(p\vqh,\vq\, )$ has simple structure
in a coordinate system, in which $z$--axis directed along the vector $\vqh.$
Considering arbitrary rotations about the $z$--axis we can see that
\be\label{PsiSMmmCollinear} \Psi^{SMm}_{m'}(p\vqh,\vq\, ) \sim\delta_{M+m,m'}. \ee
WF~\eqref{PsiSMmmCollinear} may take nonzero values for quantum numbers
\begin{align}
S&=0: & mm'&=\Oh\Oh,\:-\Oh-\!\Oh,                                 \label{S0mm}\\
S&=1: & Mmm'&=1-\!\Oh\Oh,\:0\Oh\Oh,\:0-\!\Oh-\!\Oh,\:-1\Oh-\!\Oh. \label{S1Mmm}
\end{align}
Six complex components of WF~\eqref{PsiSMmmCollinear}
with quantum numbers~\eqref{S0mm} and~\eqref{S1Mmm} satisfy six real conditions.
So, WF~\eqref{PsiSMmmCollinear} depends on six real scalar functions.

From~\eqref{PhiN0}--\eqref{PhiVecN1} 
we get
\begin{align}
S&=0: & \Phi(p\vqh,\vq\:)&=\bar\psi_1 + \vec{\sigma}\cdot\vqh\;\bar\psi_9,
\label{PhiCollin}\\                                                               
S&=1: & \vec{\Phi}(p\vqh,\vq\:)&=\vec{\sigma} \; \bar\psi_4
+ \vqh \; \vec{\sigma}\cdot\vqh \; \bar\psi_6
+\vqh\;\bar\psi_{11} + i\bigl[\vec{\sigma}\times\vqh\:\bigr]\;\bar\psi_{13},
\label{PhiVecCollin}\end{align}                                                   
where $\bar\psi_\nu=\bar\psi_\nu(p,q)$ and $\nu=1,4,6,9,11,13.$

One can express $\bar\psi_\nu$ through components of tensor WF~\eqref{PsiSMmmCollinear} via
$ \bar\psi_{\nu'}=\phi_{\nu}(p,q,1),$ where $\nu'=1,9,11,13$ corresponds to $\nu=1,2,5,15,$
and \[ \bar\psi_4=\Oh(\phi_8-\phi_9), \qquad \bar\psi_6=\Oh(-\phi_8+3\phi_9). \]
Functions $\phi_\nu(p,q,1)$ are given
    by~\eqref{phiS0} with $\nu=1,2$
and by~\eqref{phiS1} with $\nu=5,8,9,15.$

Properties of the WF for antiparallel vectors $\vp$ and $\vq$ are
the same as in the  case of parallel relative momenta treated above.


\setcounter{equation}0
\section{Relations of functions $\psi_\nu$ to components \\ of the tensor wave function}
\label{Apppsi2phi}

In this Appendix scalar functions $\psi_\nu(p,q,\xi),$
used to construct the 3N bound state WF by means of~\eqref{PhiN0}--\eqref{PhiVecN1},
are written as superpositions of $\phi_\nu(p,q,\xi).$
The latter can be related to $\Psi^{SMm}_{m'}(\vp,\vq\, )$
with the help of~\eqref{phiS0} and~\eqref{phiS1}.

Aiming to express functions  $\psi_\nu$ through $\phi_\nu,$
we insert \eqref{PhiN0},    \eqref{PhiN1}    into~\eqref{phiS0}
      and~\eqref{PhiVecN0}, \eqref{PhiVecN1} into~\eqref{phiS1}.
Matrix $U$ of linear transformation
$\phi_{\nu'}=\sum_{\nu} U_{\nu'\nu}\psi_{\nu},$ where $\nu,\nu'=1,\dots,16,$
has diagonal--block structure and is degenerated when $\vph=\pm \vqh.$
The case of collinear relative momenta is discussed in \ref{AppCollinearRelMom}
Here we consider the inverse transformation when $\xi\neq\pm1.$

We have $\psi_1=\phi_1$ and $(1-\xi^2)\psi_\nu=\phi_{\nu'},$ 
where $\nu'=4,7$ correspond to $\nu=2,3.$

Functions $\psi_\nu$    with $\nu=4,\dots,8$ depend on
          $\phi_{\nu'}$ with $\nu'=8,\dots,12$
\begin{align}
(1-\xi^2)\psi_4&=\phi_{19},
\label{psi4tophi}
\\
(1-\xi^2)^2\psi_5
&=\xi(1-\xi^2)\phi_8-2\xi(\phi_9+\phi_{10})+              \nonumber\\
&\hphantom{=}+(1+3\xi^2)\phi_{11}+\Oh(1-\xi^2)\phi_{12},
\label{psi5tophi}
\\
(1-\xi^2)^2\psi_6
&=-(1-\xi^2)\phi_8+2\phi_9+(1+\xi^2)\phi_{10}-4\xi\phi_{11},
\label{psi6tophi}
\\
(1-\xi^2)\psi_7&=(1-\xi^2)\psi_6-\phi_9+\phi_{10},
\label{psi7tophi}
\\
(1-\xi^2)\psi_8&=(1-\xi^2)\psi_5-\phi_{12},
\label{psi8tophi}
\end{align}
where $\phi_{19}$ is defined by~\eqref{phi19phi}.

Relations between the functions that determine parity--odd component of the 3N bound state
can be written as
\be\label{psi916phi} (1-\xi^2)A_\mu=G B_\mu, \quad \mu=1,\dots,4, \qquad
 G=\bigl(\begin{smallmatrix} 1&-\xi \\ -\xi&1 \end{smallmatrix}\bigr). \ee
Vectors $A_\mu, B_\mu$ are given in Table~\ref{phi2PsiAmuBmu}.

\renewcommand{\arraystretch}{1.5}

\begin{table}[h]
\begin{center}
\caption[Table]{\label{phi2PsiAmuBmu}
Vectors $A_\mu$ and $B_\mu$ used in Eqs.~\eqref{psi916phi}
that relate function $\psi_\nu$ and $\phi_{\nu'}$ with
$\nu=9,\dots,16$ and $\nu'=2,3,5,6,13,\dots,16$
}
\renewcommand{\arraystretch}{1.5}
\begin{tabular}[t]{|ccc|}                                                \hline
$\mu$ & $A_\mu$                    & $B_\nu$                          \\ \hline
1 $\vphantom{A^{\Psi^9}_{\Psi_9}}$
&
$ \bigl(\begin{smallmatrix} \psi_9   \\\psi_{10} \end{smallmatrix}\bigr) $
&
$ \bigl(\begin{smallmatrix} \phi_{2} \\\phi_{3}  \end{smallmatrix}\bigr) $
\\
2
&
$ \bigl(\begin{smallmatrix} \psi_{11}\\\psi_{12} \end{smallmatrix}\bigr) $
&
$ \bigl(\begin{smallmatrix} \phi_5   \\\phi_{6}  \end{smallmatrix}\bigr) $
\\ \hline
\end{tabular}
\hspace*{5mm}
\begin{tabular}[t]{|ccc|}                                                \hline
$\mu$ & $A_\mu$                    & $B_\nu$                          \\ \hline
3
&
$ \bigl(\begin{smallmatrix} \psi_{13}\\\psi_{14} \end{smallmatrix}\bigr) $
&
$ \bigl(\begin{smallmatrix} \phi_{15}\\\phi_{16} \end{smallmatrix}\bigr) $
\\
4
&
$ {\scriptstyle (1-\xi^2)}
  \bigl(\begin{smallmatrix} \psi_{15}\\\psi_{16} \end{smallmatrix}\bigr) $
&
$ \bigl(\begin{smallmatrix} \phi_{13}\\\phi_{14} \end{smallmatrix}\bigr) $
\\ \hline
\end{tabular}
\end{center}
\end{table}

\end{document}